\documentclass[twoside,12pt]{article}
\usepackage{amsmath,bbm,fancyhdr,a4,epsfig,umlaut,amssymb,psfrag,multirow,exscale,caption2}
\newcommand{\Section}[1]{\section{#1} \setcounter{equation}{0}}

\allowdisplaybreaks[1]
\newcounter{definition}
\newenvironment{defn}[1]
{\bigskip\begin{sloppypar}\noindent\stepcounter{definition}%
\textbf{Definition\arabic{definition}}\,({#1})\\}
{\end{sloppypar}}
\newcounter{theorem}
\newenvironment{theo}[1]
{\bigskip\begin{sloppypar}\noindent\stepcounter{theorem}%
\textbf{Theorem\arabic{theorem}}\,({#1})\\\itshape}
{\end{sloppypar}}

\begin{document}

\title{
\vskip -70pt
\begin{flushright}
{\normalsize \ DAMTP-2003-72}\\
\end{flushright}
\vskip 25pt
{\bf Delocalised Spinors}
}
\vspace{1.4cm}
{\makeatletter
\author{N. S. Manton\thanks{e-mail address: N.S.Manton@damtp.cam.ac.uk} \hspace{1pt} and  A. F. Schunck{\thanks{e-mail address: A.F.Schunck@damtp.cam.ac.uk}} \\ \small{\textsl{Department of Applied Mathematics and Theoretical
Physics}}
\\ \small{\textsl{University of Cambridge}} \\
\small{\textsl{Wilberforce Road, Cambridge CB3 0WA, England}}}
\makeatother}

\date{\today}
\maketitle

\begin{abstract}
Solutions to the four-dimensional Euclidean Weyl equation in the
background of a general JNR $N$-instanton are known to be
normalisable and regular throughout four-space. We show that these
solutions are asymptotically given by a linear combination of
simple singular solutions to the free Weyl equation, which can be
interpreted as localised spinors. The `spinorial' data
parameterising the asymptotics of the delocalised solutions to the
Weyl equation in the presence of the instanton almost determines
the background instanton, yet not completely. However, it captures
the geometry and symmetry of the underlying instanton
configuration.

\end{abstract}

\Section{Introduction} It is a well-known fact that in the case of
massless particles the Weyl equation, instead of the Dirac
equation, provides an adequate description of particles with
negative helicity, ignoring positive helicity particles, or
conversely. For negative helicity particles one finds that in
two-dimensional Euclidean space-time the Weyl equation corresponds
to the Cauchy-Riemann equations. Thus the class of solutions is
given by the holomorphic functions. To interpret the solutions as
localised classical particles, they must approach zero at
infinity. If one also insists on their analyticity in the whole of
Euclidean two-space, the only possible solutions are the constants
(Liouville's theorem). Hence, any \textsl{interesting} solutions,
that tend to zero at infinity, must exhibit a singularity. In the
simplest case these solutions are therefore just given by $f(z) =
a/(z - b)$, characterised by two complex constants, the
residue $a$ of the function and the position parameter $b$.
However, the singularity makes the interpretation of these
solutions as well-behaved particles again difficult.

In Euclidean four-space we find a similar situation. Here the Weyl
equation for right-handed particles corresponds to the so-called
Cauchy-Riemann-Fueter equation, the quaternionic analogue of the
Cauchy-Riemann equations. Just as holomorphic functions satisfy
the latter, quaternionic \textsl{regular} functions satisfy the
former. This definition of regular for quaternionic functions is
the basis of quaternionic analysis, which provides us with
quaternionic analogues of e.g. Cauchy's theorem, Cauchy's integral
formula and the Laurent expansion \cite{Sudbery}. Since the proof
of Liouville's theorem depends only on Cauchy's integral formula,
there exists also the quaternionic analogue of this theorem.
Hence, again any \textsl{interesting} regular function, that
approaches zero at infinity, has to be singular in at least one
point and thus, cannot easily be interpreted as a classical
particle.

However, if one considers the Weyl equation in the background of
an instanton, i.e. a topologically non-trivial solution to the
four-dimensional Euclidean Yang-Mills equation, one finds that the
instanton `washes out' the singularity, leaving as a solution an
everywhere nicely behaved particle. By that we mean that the
asymptotic behaviour of the solution to the free Weyl equation and
the solution in the instanton background can be shown to agree to
appropriate order, i.e. up to and including terms of
$\mathcal{O}(r^{-4})$. Since the solution to the free equation and
the `instanton solution' are topologically inequivalent, a
comparison of the two solutions depends on a careful choice of
gauge. It can be seen from the quaternionic Laurent series, that
the first term of the expansion, which is of order
$\mathcal{O}(r^{-3})$, is similar to the familiar dipole term in
three space dimensions, whereas the second is the analogue of a
quadrupole term. The dipole term is thus characterised by one
quaternionic constant, which is the quaternionic analogue of a
residue. In contrast the quadrupole term is characterised by three
quaternionic constants, the interpretation of which is less clear.
However, they can be shown to reflect the symmetry of the
underlying instanton configuration.

In the next section we will briefly explain the concept of an
\textsl{instanton} \cite{Rajaraman}, followed by the description of a
special class of instantons, the so called Jackiw-Nohl-Rebbi
(JNR) instantons \cite{JNR}. We then go on to describe the
solutions to the four-dimensional Euclidean Dirac equation in the
background field of an arbitrary JNR instanton, as investigated by
Grossman \cite{Grossman}. Before we start describing our own
results we will then briefly introduce the theorems of
quaternionic analysis \cite{Sudbery}, that are most important for
us.

\Section{Euclidean Yang-Mills configurations} In four-dimensional
Euclidean space-time the Yang-Mills field equations possess
localised solutions having a finite Euclidean action. These
solutions are referred to as \slshape{instantons}. \upshape
Topologically distinct from the trivial absolute minimum of the
action functional, instantons do not have a vanishing field
strength $G_{\mu \nu}$. However, $G_{\mu \nu}$ has to vanish on
the boundary of Euclidean four-space, $S^{3}_{\infty}$, which
implies that the corresponding fields $A_{\mu}$ tend
asymptotically to a pure gauge, thus defining a mapping from the
surface at infinity into the group space of the gauge group.
Instantons are therefore characterised by a topological invariant
$N$, the so-called Pontryagin index, which takes integer values.

In the case of an SU(2) non-Abelian gauge theory this index is
essentially the `winding number' of $S^{3}_\infty$ into the group
space of SU(2) $\cong S^3$. Therefore it is equal to the homotopy
index, which characterises the discrete infinity of homotopy
classes of the mapping $S^3 \rightarrow S^3$. Within each class
$N$, the action is bounded from below by a constant multiple of
$|N|$ and the absolute minimum value is in each class attained
when the field strength satisfies $G_{\mu \nu} = \pm *G_{\mu
\nu}$, where $*G_{\mu \nu}$ is the dual field tensor given by
$*G_{\mu \nu} = \frac{1}{2} \epsilon_{\mu\nu\rho\sigma} G_{\rho
\sigma}$.

Strictly speaking, the notion \slshape instanton\upshape\ only
refers to self-dual solutions of the Yang-Mills equation whereas
anti-self-dual solutions are referred to as \slshape anti-\upshape
instan\-tons. The fundamental difference between instantons and
anti-instantons is given by the sign of the Pontryagin index,
which is positive for instantons but negative for anti-instantons.
For definiteness we will in the following only consider instantons
though most results will be applicable to anti-instantons as well.

\Section{JNR instantons}
 The general $N$-instanton solution has been constructed implicitly
 by Atiyah et al. in 1978 \cite{AHDM} for an arbitrary compact classical
group. In the case of SU(2) as the gauge group, $8N -3$ parameters
characterise the solutions. A subset of these solutions can be
constructed explicitly. These special instantons exhibit only $5N
+ 4$ parameters and were investigated by Jackiw, Nohl and Rebbi
already in 1977 \cite{JNR}. However, in the case of an $N=2$
instanton this JNR solution to the self-duality equation is the
most general one since an $N=2$ instanton is fully described by 13
physical parameters. The fourteenth parameter occurring in the JNR
formula corresponds to a gauge transformation. An even more
restrictive case is the one of a 't Hooft instanton \cite{'tHooft}
which is characterised by $5N$ parameters, thus giving the most
general $N=1$ instanton. The $N=1$ 't Hooft instanton is related
to the JNR $N=1$ instanton via a gauge transformation
\cite{Manton}.

\bigskip

We begin by briefly summarising the construction of the JNR
$N$-instanton solution as described in \cite{JNR, Rajaraman}. To
simplify notation we will consider a matrix representation of the
three vector fields $A_\mu^a$, $a = 1,2,3$, taking values in the
Lie Algebra of SU(2)
\begin{equation}
\label{SU(2)vectorfield} A_\mu = A_\mu^a \frac{\sigma_a}{2i}.
\end{equation}
Here $\sigma_a$ are the familiar Pauli matrices, the three
generators of the group SU(2). The field strength is given by
\begin{equation}
G_{\mu \nu} = \partial_\mu A_\nu - \partial_\nu A_\mu + [A_\mu,
A_\nu].
\end{equation}
It is useful to define the following matrices
\begin{equation}
\alpha_0 = \mathbbm{1}_2, \quad {\alpha}_i = - i \sigma_i, \quad
\bar{\alpha}_0 = \mathbbm{1}_2, \quad \bar{\alpha}_i = i \sigma_i
\end{equation}
and
\begin{equation}
\bar{\sigma}_{\mu\nu} = \frac{1}{4i}\left(
\bar{\alpha}_\mu\alpha_\nu
 - \bar{\alpha}_\nu \alpha_\mu \right).
\end{equation}
Notice that $\bar{\sigma}_{\mu\nu}$ is both antisymmetric and
anti-self-dual in its indices. In order to solve the self-duality
equation for the field strength $G_{\mu \nu}$
\begin{equation}
G_{\mu \nu} = *G_{\mu \nu}
\end{equation}
one makes an ansatz of the form
\begin{equation}
\label{vectorfield1} A_{\mu} = i\bar{\sigma}_{\mu\nu}a_\nu \quad
{\rm{with}} \quad a_\nu =
\partial_{\nu}\ln\rho
\end{equation}
where $\rho(\mathbf{x})$ is a scalar potential to be obtained,
which must satisfy $\triangle \rho(\mathbf{x}) = 0$. Here
$\triangle$ is the four-dimensional Euclidean Laplace operator.
Notice that the anti-self-dual symbols $\bar{\sigma}_{\mu\nu}$ in
the fields $A_\mu$ lead to a self-dual field strength. One finds
for $\rho(\mathbf{x})$\footnote{'t Hooft's solution takes the form
$\rho(\mathbf{x})= 1 + \sum_{k=1}^{N}
\frac{\lambda_{(k)}^2}{(\mathbf{x} - \mathbf{x}_{(k)})^2}$ }
\begin{equation}
\label{rho} \rho(\mathbf{x}) = \sum_{k=1}^{N + 1}
\frac{\lambda_{(k)}^2}{(\mathbf{x} - \mathbf{x}_{(k)})^2}.
\end{equation}
This solution when inserted in Eqn.~\/(\ref{vectorfield1}) will
yield the $N$-instanton solution with homotopy index $N$. Note
that although $\rho$ has a singularity at each point
$\mathbf{x}_{(k)}$, the instanton field has no singularity.

It is important to note that the scalar potential $\rho$ depends
on $N + 1$ position parameters $\mathbf{x}_{(k)}$. Thus there is
no simple relation between these parameters and the positions of
the $N$ instantons. In contrast, a 't Hooft instanton has a
potential depending on $N$ position parameters indeed
corresponding to the positions of the instantons. The other $N$
JNR parameters --- a common rescaling of the $\lambda_{(k)}$'s
does not affect the fields $A_\mu$ --- the weights, correspond to
the size of the instantons.

As shown by Jackiw, Nohl and Rebbi \cite{JNR} the ansatz in
Eqn.~\/(\ref{vectorfield1}) completely fixes the gauge, as long as
the $N$ points $\mathbf{x}_{(k)}$ do not lie on a circle (or on a
straight line, which is a circle through the point at infinity).
However, three points always lie on a circle and hence in the case
of an $N = 2$ instanton one of the 14 parameters corresponds to a
gauge transformation which moves the $\mathbf{x}_{(k)}$ around the
circle.

\Section{Solutions to the massless Dirac equation in an
$N$-instanton background} Solutions to the massless Dirac equation
in the background of an $N$-instanton  have been explicitly
constructed by Grossman \cite{Grossman} with the gauge group being
SU(2). Later on Corrigan et al. \cite{Corrigan} and independently
Osborn \cite{Osborn} derived these solutions for an SU($n$) or
Sp($k$) gauge group, respectively. However the latter solutions
are only implicitly known. In the following we will briefly
explain Grossman's construction of which we shall make extensive
use in due course\footnote{For later convenience we will, however,
use a slightly different notation.}.

\bigskip

The zero-modes of the massless Dirac equation can be chosen to be
chiral eigenfunctions of $\gamma_5$. According to Grossman it is
possible to construct $N$ zero-energy modes of negative helicity,
when the field strength is chosen to be
self-dual\footnote{Grossman's actual derivation holds for positive
helicity particles. However, as pointed out by Grossman himself,
the analysis will likewise lead to $N$ negative helicity solutions
when considering self-dual fields $G_{\mu\nu}$, rather than
anti-self-dual fields.}. A version of the Atiyah-Singer index
theorem for isospinor fermion fields asserts that there are no
solutions of positive helicity, since the difference between
negative and positive helicity solutions must equal in absolute
value the Pontryagin index $N$.

The covariant massless Dirac equation in the background of an
SU(2) vector potential $A_\mu = A_\mu^a (\sigma_a/2i)$ is given by
\begin{equation}
(\partial_\mu + A_\mu)\gamma_\mu \psi = 0 .
\end{equation}
We will choose a representation of the $\gamma$ matrices in which
the equations for positive and negative helicity fields decouple,
i.e. a representation in which $\gamma_5 =
\gamma_0\gamma_1\gamma_2\gamma_3 $ is diagonal. This is given by
\begin{equation}
\gamma_0 = \left( \!\!
\begin{array}{cc}
  0 & \mathbbm{1}_2 \\
  \mathbbm{1}_2 & 0 \\
\end{array} \!\!
\right), \quad \gamma_i = \left( \!\!
\begin{array}{cc}
  0 & -i\sigma_i \\
  i\sigma_i & 0 \\
\end{array} \!\!
\right), \quad \gamma_5 = \left( \!\!
\begin{array}{cc}
  - \mathbbm{1}_2 & 0 \\
  0 & \mathbbm{1}_2\\
\end{array} \!\!
\right).
\end{equation}
Setting $\psi = \left(\!\!
\begin{array}{c}
  \psi_R \\
  \psi_L \\
\end{array}\!\!
\right)$ such that $\psi_R$ and $\psi_L$ are $2 \times 2$ matrices
in spin and isospin we get decoupled equations for the
right-handed and left-handed spinor:
\begin{eqnarray}
\label{backgroundWeyl2}
(\partial_\mu + A_\mu)\bar{\alpha}_\mu\psi_R & = & 0 \\
\label{backgroundWeyl1} (\partial_\mu + A_\mu)\alpha_\mu\psi_L & =
&  0.
\end{eqnarray}
It will be convenient to slightly rewrite
Eqn.~\/(\ref{backgroundWeyl2}). We have, preserving all indices
for the moment,
\begin{equation}
\label{backgroundWeyl2ind} \left(\partial_\mu \delta_{ij} +
(A_\mu)_{ij}\right)(\bar{\alpha}_\mu)_{\beta\gamma} \psi_{\gamma
j} = 0.
\end{equation}
Using the fact that $\sigma_k^{\rm{t}} = \epsilon \sigma_k
\epsilon$, where $\epsilon$ is the usual totally antisymmetric
tensor with $\epsilon_{12} = 1$, and redefining the field
$\psi_R'= \psi^{\rm{t}}_R \epsilon$,  such that $\psi_R'$ is a $2
\times 2$ matrix in isospin and spin, with components $\psi'^R_{j
\alpha}$ (here $j = 1,2$ is the isotopic index and $\alpha = 1,2$
is the Lorentz index), Eqn.~\/(\ref{backgroundWeyl2ind}) becomes
\begin{equation}
\left(\partial_\mu \delta_{ij} +
(A_\mu)_{ij}\right)\psi_{j\beta}'(\alpha_\mu)_{\beta\gamma} = 0,
\end{equation}
or in index-free notation
\begin{equation}
\label{backgroundWeyl} (\partial_\mu + A_\mu)\psi_R' \alpha_\mu =
0.
\end{equation}
Note that all products occurring in Eqn.~\/(\ref{backgroundWeyl})
are now to be interpreted as matrix products. Defining
\begin{equation}
\label{Phik}
\phi^{(k)} = \frac{\lambda_{(k)}^2}{(\mathbf{x} -
\mathbf{x}_{(k)})^2}, \quad k = 1,\ldots, N + 1
\end{equation}
and
\begin{equation}
\label{Muk}
M_\mu^{(k)} = \rho^{1/2}\partial_\mu \left(
\frac{\phi^{(k)}}{\rho}\right),
\end{equation}
with $\rho$ given by Eqn.~\/(\ref{rho}), one finds $N + 1$
solutions of the form
\begin{equation}
\label{righthandedsolution} {\psi'}_R^{(k)} =  M_\beta^{(k)}
\bar{\alpha}_\beta.
\end{equation}
However, since
\begin{equation}
\sum_{k=1}^{N + 1} M_\mu^{(k)} = 0
\end{equation}
there are only $N$ linearly independent solutions.

\bigskip

In the following we will briefly describe some major results of
the theory of quaternionic regular functions, as they characterise
the class of solutions to the four-dimensional Euclidean Dirac
equation. In our description we shall follow the lines of
Sudbery's work \cite{Sudbery}, which gives a self-contained and
rigorous account of the theory of quaternionic analysis.

\Section{Quaternionic analysis} The theory of quaternionic
analysis was developed by Fueter and his collaborators in the
years following 1935, when Fueter proposed the definition of
\slshape regular \upshape\ for quaternionic functions
\cite{Fueter}. By means of an analogue of the Cauchy-Riemann
equations this led to a theory of regular functions similar to the
theory of holomorphic functions.

\bigskip

A generic quaternion can be written as
\begin{equation}
\label{quaternion} q = 1x_0 + ix_1 +  jx_2 + kx_3 \quad
\mathrm{with} \quad x_\mu \in \mathbbm{R}, \quad\mu = 0,\ldots,3,
\end{equation}
where $1$, $i$, $j$ and $k$ shall denote the elements of the
standard basis for $\mathbbm{R}^4$. One defines the quaternionic
product on $\mathbbm{R}^4$ as the $\mathbbm{R}$-bilinear product
\begin{equation}
\mathbbm{R}^4 \times \mathbbm{R}^4 \rightarrow \mathbbm{R}^4;\quad
(q_1,q_2)\mapsto q_1q_2
\end{equation}
with unit element 1, such that
\begin{equation}
i^2 = j^2 = k^2 = -1
\end{equation}
\begin{equation}
i j = k = -ji, \quad jk = i = -kj, \quad ki=j=-ik.
\end{equation}
Here we will identify the subfield spanned by 1 with
$\mathbbm{R}$. Notice that the quaternionic product is not
commutative. The linear space $\mathbbm{R}^4$ with the
quaternionic product defines the real associative algebra
$\mathbbm{H}$ of the quaternions (see, for example,
\cite{Porteous}).

We will sometimes use $e_i$, $i = 1, 2, 3$, to denote the basic
quaternions $i$, $j$ and $k$. Then Eqn.~\/(\ref{quaternion})
becomes
\begin{equation}
q = x_0 + e_ix_i,
\end{equation}
or setting $e_0 = 1$
\begin{equation}
q = e_\mu x_\mu,
\end{equation}
where summation over repeated indices is implied.

Identifying the sub\-field spanned by 1 and $k$ with the complex
field $\mathbbm{C}$ we will sometimes write
\begin{equation}
\label{complexq} q = y + jz
\end{equation}
where $y= x_0 + kx_3$ and $z= x_2 + kx_1$. The conjugate of $q$,
denoted by $\bar{q}$, is given by
\begin{equation}
\bar{q} = x_0 - ix_1 - jx_2 - kx_3
\end{equation}
and the modulus by
\begin{equation}
|q| = \sqrt{q\bar{q}} = \sqrt{x_0^2 + x_1^2 + x_2^2 + x_3^2} \quad
\in \mathbbm{R}.
\end{equation} We have
\begin{eqnarray}
\overline{q_1q}_2 & = & \bar{q}_2\bar{q}_1 \quad \rm{and}\\
q^{-1} & = & \frac{\bar{q}}{|q|^2}.
\end{eqnarray}

The following definition of a \textsl{regular} function is the
most convenient for our purposes (Sudbery gives a different
definition, but shows that it is equivalent to this).
\begin{defn}{the Cauchy-Riemann-Fueter equations}
A real-differentiable, quaternion-valued function $f :
\mathbbm{R}^4 \rightarrow \mathbbm{H}$ is right-regular at $q$ if
and only if
\begin{equation}
\label{CRFequation} \frac{\partial f}{\partial x_0} +
\frac{\partial f}{\partial x_1} i + \frac{\partial f}{\partial
x_2} j + \frac{\partial f}{\partial x_3} k = 0.
\end{equation}
It is left-regular at $q$ if and only if
\begin{equation}
\frac{\partial f}{\partial x_0} + i \frac{\partial f}{\partial
x_1} + j \frac{\partial f}{\partial x_2} + k \frac{\partial
f}{\partial x_3} = 0.
\end{equation}
\end{defn}
The analysis of regular quaternionic functions is comparable to
that of holomorphic functions, since the quaternionic analogues of
Cauchy's theorem, Cauchy's integral theorem and the Laurent series
exist for regular functions. As many of the standard theorems of
complex analysis depend only on Cauchy's theorem, they also hold
for quaternionic regular functions. An example of this is
Liouville's theorem. Hence, a quaternionic function, which is
regular in the whole of $\mathbbm{R}^4$ and that tends to zero as
$|q| \rightarrow \infty$ must necessarily reduce to a constant.

We will in the following consider only right-regular functions
which we shall call simply \textsl{regular}\footnote{Sudbery
chooses to develop the theory of quaternionic analysis for
left-regular functions. However, his results are easily rewritten
for right-regular functions.}.

We may introduce the following notation for the differential
operators
\begin{eqnarray}
\label{rightreg} \bar{\partial}_r f & = &
\frac{1}{2}\left(\frac{\partial
f}{\partial x_0} + \frac{\partial f}{\partial x_i} \, e_i   \right)\\
\partial_r f & = & \frac{1}{2}\left(\frac{\partial
f}{\partial x_0} - \frac{\partial f}{\partial x_i}\, e_i \right)\\
\bar{\partial}_l f & = & \frac{1}{2}\left(\frac{\partial
f}{\partial x_0} + e_i \, \frac{\partial f}{\partial x_i}  \right)\\
\partial_l f & = & \frac{1}{2}\left(\frac{\partial
f}{\partial x_0} - e_i\, \frac{\partial f}{\partial x_i}  \right).
\end{eqnarray}
Notice that $\bar{\partial}_r$, $\partial_r $, $\bar{\partial}_l$
and $\partial_l$ all commute and that we have
\begin{equation}
\label{diffop} \triangle = 4 \,  \partial_r \, \bar{\partial}_r =
4 \, \partial_l \, \bar{\partial}_l.
\end{equation}
Thus, a function $f$ is regular if and only if $\bar{\partial}_r f
= 0$.

As shown by Sudbery all regular functions are necessarily
infinitely differentiable. It follows then from the above theorem,
together with Eqn.~\/(\ref{diffop}), that all regular functions
are harmonic.

Setting $q = y + jz$, as in Eqn.~\/(\ref{complexq}), and $f= h +
gj$, where $g$ and $h$ are complex-valued functions, we find that
Eqn.~\/(\ref{CRFequation}) is equivalent to the pair of complex equations
\begin{eqnarray}
\label{complexifiedCRF}
-\partial_{\bar{z}} g + \partial_{\bar{y}} h & = & 0 \nonumber \\
\partial_{y} g + \partial_{z} h & = & 0.
\end{eqnarray}
In the absence of a background field these complex
equations correspond to Eqn.~\/(\ref{backgroundWeyl}), which is
the Weyl equation for right-handed spinors. Note that it is
permitted to multiply $f$ with an arbitrary constant from the
left. Thus $-jf = \bar{g} - \bar{h}j$ will be a solution to
Eqn.~\/(\ref{CRFequation}), too.
\bigskip

For our purposes the last section of Sudbery's paper, which
addresses the theory of regular power series, is especially
important. Sudbery introduces the following differential operator
\begin{equation}
\partial_\nu = \frac{\partial^n}{\partial x_{i_1} \ldots \partial x_{i_n} } =
\frac{\partial^n}{\partial {x_1}^{n_1}\partial {x_2}^{n_2}\partial
{x_3}^{n_3}}.
\end{equation}
This calls for some explanation. $\nu$ is an unordered set of $n$
integers $\{i_1,\ldots,i_n\}$ with $1 \le i_r \le 3$. The number
of 1's in $\nu$ is given by $n_1$, the number of 2's by $n_2$ and
the number of 3's by $n_3$ such that $n_1 + n_2 + n_3 = n$. Hence
there are $\frac{1}{2}(n + 1)(n + 2)$ such sets $\nu$ and we will
denote the collection of these sets as $s_n$. They are to be used
as labels. If $n = 0$, i.e. $\nu = \emptyset$, we use the suffix 0
instead of $\emptyset$.

We will furthermore need the following two functions:
\begin{equation}
\label{Gq} G_\nu (q) = \partial_\nu \, G(q) \quad \mbox{with}
\quad G_0(q) \equiv G(q) = \frac{q^{-1}}{|q|^2}
\end{equation}
and
\begin{equation}
P_\nu (q) = \frac{1}{n!} \sum \, (x_0 e_{i_1} - x_{i_1})\ldots
(x_0 e_{i_n} - x_{i_n}) \quad \mbox{with} \quad P_0(q) = 1
\end{equation}
where the sum is over all $n!/(n_1! n_2! n_3!)$ different
orderings of $n_1$ 1's, $n_2$ 2's and $n_3$ 3's. Note that the
$P_\nu$ are both right- and left-regular. We have
\begin{equation}
\frac{1}{2\pi^2}\int_S  P_\mu(q) D\!q G_\nu(q) = \delta_{\mu \nu}
\end{equation}
and, since the $P_\nu$ are also left-regular, we have as well
\begin{equation}
\frac{1}{2\pi^2}\int_S G_\mu(q) D\!q P_\nu(q) = \delta_{\mu \nu}.
\end{equation}
Here $S$ is any three-sphere surrounding the origin and the
measure $D\!q$ is the quaternion valued three form
\begin{equation}
D\!q = dx_1 \wedge dx_2 \wedge dx_3 - \epsilon_{ijk} \, e_i \,
dx_0 \wedge dx_j \wedge dx_k.
\end{equation}
Sudbery then proves the following theorem:
\begin{theo}{the Laurent series}
Suppose f is regular in an open set U except possibly at $q_0 \in
U$. Then there is a neighbourhood N of $q_0$ such that if $q \in
N$ and $q \ne q_0$, $f(q)$ can be represented by a series
\begin{equation}
\label{Laurent} f(q) = \sum_{n = 0}^\infty \, \sum_{\nu \, \in \,
s_n} \left( a_\nu P_\nu (q - q_0) + b_\nu G_\nu (q - q_0) \right)
\end{equation}
which converges uniformly in any hollow ball
\begin{displaymath}
\{q: r \le |q - q_0| \le R\}, \quad with \quad r > 0, \quad
\mbox{which lies inside N}.
\end{displaymath}
The coefficients $a_\nu$ and $b_\nu$ are given by
\begin{eqnarray}
a_\nu & = & \frac{1}{2\pi^2} \int_C  f(q) D\!q G_\nu(q - q_0)  \\
\label{Laurentseriesbs} b_\nu & = & \frac{1}{2\pi^2} \int_C  f(q)
D\!q P_\nu(q - q_0) ,
\end{eqnarray}
where C is any closed 3-chain in $U - \{q_0\}$ which is homologous
to $\partial B$ for some ball $B$ with $q_0 \in B \subset U$ (so
that C has wrapping number 1 about $q_0$).\footnote{For the exact
definition of the wrapping number see \cite{Sudbery}.}
\end{theo}

\Section{The Dirac Equation in a quaternionic notation} As
remarked in the previous section, the Weyl equation for
right-handed spinors in the absence of an external
field\footnote{We will drop the subscript $R$ from now on.}
\begin{equation}
\label{Weyl} \partial_\mu \psi'\alpha_\mu =  0
\end{equation}
corresponds to the complex version of the Cauchy-Riemann-Fueter
equation, when using the complex variables $y = x_0 + i x_3$, $z =
x_2 + i x_1$
\begin{eqnarray*}
-\partial_{\bar{z}} \psi'_1 + \partial_{\bar{y}} \psi'_2 & = & 0 \nonumber \\
\partial_{y} \psi'_1 + \partial_{z} \psi'_2 & = & 0.
\end{eqnarray*}
Here  $\left( \psi'_1, \psi'_2 \right)$ are the two complex
components of $\psi'$.

The correspondence of the Cauchy-Riemann-Fueter equation to the
free Weyl equation can be seen more directly. Since the group of
quaternions with absolute value 1, namely the symplectic group
Sp(1), is isomorphic to $S^3$ and thus to SU(2) we can rewrite
Eqn.~\/(\ref{Weyl}) identifying $-i \sigma_1$ with the
quaternionic $i$, $-i \sigma_2$ with $j$ and $-i \sigma_3$ with
$k$. This yields directly the Cauchy-Riemann-Fueter equation,
Eqn.~\/(\ref{CRFequation}), in terms of quaternionic variables
\begin{equation}
(\partial_0 +  \partial_1 i_r + \partial_2 j_r +  \partial_3 k_r)
\psi'_q = 0.
\end{equation}
Here the subscript \textsl{r} shall denote action from the right.
Note that $\psi'_q$ is a single quaternionic function. It is
readily seen that $\psi'_q$ is in terms of the complex functions
$\psi'_1$ and $\psi'_2$ given by
\begin{equation}
\psi'_q = \psi'_1 j + \psi'_2.
\end{equation}

\bigskip

We now turn to the discussion of the Dirac equation in the
background field of an SU(2) vector potential given by $A_\mu =
-\frac{i}{2} A_\mu^a \sigma_a$. The Weyl equation for negative
helicity particles is given by Eqn.~\/(\ref{backgroundWeyl})
\begin{equation}
\label{BackgroundWeyl} (\partial_\mu + A_\mu)\psi' \alpha_\mu = 0,
\end{equation}
where $\psi'$ is now a $2 \times 2$ matrix in isospin and spin.
Here, again, we can use the isomorphism Sp(1) $\cong$ SU(2) to
rewrite Eqn.~\/(\ref{BackgroundWeyl}). Applying this to both the
Lorentz group and the gauge group will not be possible in general,
since it would result in combining the four complex components of
$\psi'$ into a single quaternionic function. Thus it would lead to
a reduction of the four complex parameters to two complex
parameters. However, for our purpose it will be convenient to
impose the following SU(2) gauge invariant conditions on the four
complex components of $\psi'$, which will allow us to rewrite
Eqn.~\/(\ref{BackgroundWeyl}) in a purely quaternionic notation:
\begin{equation}
\label{realitycond}
\psi' = \left( \begin{array}{cc} \psi_{11} & \psi_{12} \\
\psi_{21} & \psi_{22}
 \end{array} \right) = \left( \begin{array}{cc} \psi_{11} & \psi_{12} \\
- \bar{\psi}_{12} & \bar{\psi}_{11}
 \end{array} \right).
\end{equation}
Note that these conditions can be viewed as a Majorana condition.
If we set
\begin{equation}
\Psi_M = \left(\!\!\! \begin{array}{r} \psi_{11} \\ \psi_{12} \\ -
\bar{\psi}_{12} \\ \bar{\psi}_{11}
\end{array} \!\! \right)
\end{equation}
we find
\begin{equation}
\Psi_M = \Psi_M^C \quad \mbox{with} \quad \Psi_M^C \equiv C
\overline{\Psi}_M^{\mathrm{t}}.
\end{equation}
Here $C$ is the charge conjugation operator given by $C = \gamma_0
\gamma_2$ and $\overline{\Psi} = \Psi^{\dagger}\gamma_0$ is the
usual Dirac conjugate spinor.

Setting $A_\mu^q = \frac{1}{2} e_a A_\mu^a$ we find for
Eqn.~\/(\ref{BackgroundWeyl})
\begin{equation}
 (\partial_\mu + A_\mu^q)\psi'_q e_\mu = 0,
\end{equation}
where $\psi'_q$ is now a single quaternionic object, given by
\begin{equation}
\psi'_q = \bar{\psi}_{11} - \bar{\psi}_{12} j = \psi_{22} +
\psi_{21} j.
\end{equation}
For the gauge transformation of $\psi'_q$ and $A_\mu^q$ we have
\begin{eqnarray}
\psi'_q  & \longrightarrow & q \psi'_q \\
A_\mu^q & \longrightarrow & q A_\mu^q q^{-1} - (\partial_\mu
q)q^{-1},
\end{eqnarray}
with $q$ a unit quaternion, i.e. $|q| = 1$.

\bigskip

As the simplest singular solution to the (complex) Eqn.~\/(\ref{BackgroundWeyl}) in
the absence of a background field we find
\begin{equation}
\psi' = \frac{1}{(y\bar{y} + z\bar{z})^2}  \left( \begin{array}{cc} \bar{a}y + b\bar{z}& \bar{a}z - b\bar{y} \\
- a\bar{z} + \bar{b}y & a\bar{y} + \bar{b}z
\end{array} \right),
\end{equation}
where we have already imposed the conditions Eqn.\/(\ref{realitycond})
on the four complex
components of $\psi'$. If
we write $a = c_0 + ic_3$, $b = c_2 + ic_1$, where we assume
$c_\mu$ to be the components of some real vector $\mathbf{c}$, we
may write instead
\begin{equation}
\label{matrixsingularsolution} \psi' =
\frac{(\mathbf{c}.{\mbox{\boldmath$\alpha$}})
(\mathbf{x}.{\bar{\mbox{\boldmath$\alpha$}}})}{r^4},
\end{equation}
with $r = \sqrt{\mathbf{x}^2}$ the distance from the origin.

In quaternionic notation we have (identifying as usual the
quaternionic $k$ with the complex $i$)
\begin{equation}
\psi'_q = \frac{1}{(y\bar{y} + z\bar{z})^2}(a + jb)(\bar{y} -
\bar{z}j).
\end{equation}
Setting $q = y + jz$ as in Eqn.~\/(\ref{complexq}) and defining
$c_q = a + jb$ we have
\begin{equation}
\label{singularsolution} \psi'_q = c_q\frac{\bar{q}}{|q|^4} = c_q
\frac{q^{-1}}{|q|^2} = c_q G(q),
\end{equation}
with $G(q)$ as defined in Eqn.~\/(\ref{Gq}). Note that this is the
general `dipole' term of the quaternionic Laurent expansion,
Eqn.~\/(\ref{Laurent}). Thus, $\psi'_q$ is regular except at the
origin.

As we shall show in the following, the solutions to
Eqn.~\/(\ref{BackgroundWeyl}) in the presence of a JNR $N$
instanton can be written up to quadrupole order, i.e.~up to and
including terms of $\mathcal{O}(r^{-4})$, as a linear combination
of shifted dipoles. However, this depends on a careful choice of
gauge, as will be shown below.

\Section{$1/r$ expansion of Grossman's solution}

In the case of a JNR $N$ instanton as the background field there
exist $N + 1$ (linearly dependent) solutions to the Weyl equation
for right-handed spinors, which are given
by Eqns.~\/(\ref{Phik}) -- (\ref{righthandedsolution})
\begin{displaymath}
 \psi'_{(k)} =  M_\beta^{(k)}
\bar{\alpha}_\beta, \quad k = 1,\ldots, N + 1.
\end{displaymath} Calculating these explicitly, we find for $M_\mu^{(k)}$
\begin{eqnarray}
\label{Grossmansolution} \lefteqn{M_\mu^{(k)}  =   \frac{-2
\lambda_{(k)}^2}{\rho^{3/2} \prod_{i=1}^{N + 1} (\mathbf{x} -
\mathbf{x}_{(i)})^4}\, \cdot}\\
&&\left[ \sum_{i \ne k} \left(\prod_{j \ne i,k} (\mathbf{x} -
\mathbf{x}_{(j)})^4 \right) \lambda_{(i)}^2 \left( (x_\mu -
x_{{(k)}\mu}) (\mathbf{x} - \mathbf{x}_{(i)})^2 - (x_\mu -
x_{{(i)}\mu})(\mathbf{x} - \mathbf{x}_{(k)})^2 \right) \right]
\nonumber,
\end{eqnarray}
with $\rho(\mathbf{x})$ given as before by
\begin{displaymath}
\rho(\mathbf{x}) = \sum_{i=1}^{N + 1}
\frac{\lambda_{(i)}^2}{(\mathbf{x} - \mathbf{x}_{(i)})^2}.
\end{displaymath}
In order to be able to compare this solution to the solution of
the free Weyl equation discussed in the previous section, one has
to investigate its asymptotic behaviour, i.e.~its behaviour far
away from the instanton. Thus we have to expand $M_\mu^{(k)}$ in
powers of $1/r$. We shall write $M_\mu^{(k)}$ as the sum of a
`dipole' and a `quadrupole term', $m_{(k)\mu}^{D}$ and
$m_{(k)\mu}^{Q}$ respectively, plus higher order contributions. It
turns out that the first term of this expansion, the dipole term,
is of $\mathcal{O}(r^{-3})$, the quadrupole term of
$\mathcal{O}(r^{-4})$. Therefore
\begin{equation}
M_\mu^{(k)} = m_{(k)\mu}^{D} + m_{(k)\mu}^{Q}+
\mathcal{O}(\frac{1}{r^5}).
\end{equation}
Defining $\Lambda$ as the sum of the weights $\lambda_{(i)}^2$ and
$\mathbf{X}$ as the `centre of mass' of the instanton
\begin{equation}
\Lambda = \sum_{i=1} \lambda_{(i)}^2 \quad \mbox{and} \quad
\mathbf{X} = \frac{\sum_{i=1} \lambda_{(i)}^2
\mathbf{x}_{(i)}}{\Lambda}
\end{equation}
the dipole contribution is given by
\begin{equation}
\label{ungaugeddipole} m_{(k)\mu}^{D} = - \frac{1}{r^3} \cdot
\frac{2 \lambda_{(k)}^2}{\Lambda^{1/2}} \cdot \left[(X_\mu -
x_{(k)\mu}) - 2(\mathbf{\hat{x}}.\mathbf{X} -
\mathbf{\hat{x}}.\mathbf{x}_{(k)}){\hat{x}}_\mu \right].
\end{equation}
If we define
\begin{equation}
\label{nhat} \hat{n}_\mu^{(i)} = 2
\hat{\mathbf{x}}.\mathbf{x}_{(i)}\hat{x}_\mu - x_{(i)\mu} \quad
\mbox{and similarly} \quad \hat{n}_\mu^{(X)} = 2
\hat{\mathbf{x}}.\mathbf{X}\,\hat{x}_\mu - X_\mu
\end{equation}
we can rewrite Eqn.~\/(\ref{ungaugeddipole}) as
\begin{equation}
m_{(k)\mu}^{D} = - \frac{1}{r^3} \cdot \frac{2
\lambda_{(k)}^2}{\Lambda^{1/2}} \cdot \left[\hat{n}_\mu^{(k)} -
\hat{n}_\mu^{(X)} \right].
\end{equation}
We find for the quadrupole term
\begin{eqnarray}
\lefteqn{m_{(k)\mu}^{Q} = - \frac{1}{r^4} \cdot \frac{2
\lambda_{(k)}^2}{\Lambda^{1/2}} \Bigg[ \hat{x}_\mu
\left(\frac{\sum_{i}\lambda_{(i)}^2 \mathbf{x}_{(i)}^2}{\Lambda} -
\mathbf{x}_{(k)}^2\right) +
2\left(\mathbf{\hat{x}}.\mathbf{X}\hat{n}_\mu^{(k)} -
\mathbf{\hat{x}}.\mathbf{x}_{(k)}\hat{n}_\mu^{(X)}\right)}
\nonumber
\\
&& {} + 3\mathbf{\hat{x}}.\mathbf{X}(\hat{n}_\mu^{(X)} -
\hat{n}_\mu^{(k)}) +
4\left(\mathbf{\hat{x}}.\mathbf{x}_{(k)}\hat{n}_\mu^{(k)} -
\frac{\sum_{i}\lambda_{(i)}^2\mathbf{\hat{x}}.\mathbf{x}_{(i)}\hat{n}_\mu^{(i)}}{\Lambda}
\right)\Bigg].
\end{eqnarray}

If one now wants to compare this solution asymptotically to the
solution of the Weyl equation in the absence of an instanton, one
encounters the difficulty that the fields $A_\mu$ tend
asymptotically to a pure gauge rather than to the vacuum, which is
topologically different.

It would be desirable to find a gauge in which the fields $A_\mu$
tend to zero at infinity more rapidly than the pure gauge does.
Because of the topologically different nature of the vacuum and
the instanton, there exists no gauge transformation that is
non-singular at all $\mathbf{x}$, relating the vacuum and the
instanton. However, it turns out that one can find a
\textsl{singular} gauge such that $A_\mu$ is (at most) of
$\mathcal{O}(r^{-3})$. An easy estimate then shows that in this
singular gauge $\psi'_{(k)}$ will be --- up to and including
quadrupole order --- not only a solution to the Weyl equation in
the instanton background but also to the free Weyl equation. Note
that since the free Weyl equation is homogeneous, i.e. does not
mix powers of $r$, the dipole term and the quadrupole term will be
independently solutions to the free equation. The next task will
obviously be to find the desired gauge transformations.

\Section{Singular gauge transformation}

Under a gauge transformation, $A_\mu$ transforms as usual as
\begin{equation}
A_\mu \longrightarrow A'_\mu = UA_\mu U^{-1} - (\partial_\mu
U)U^{-1}.
\end{equation}
Expanding $A_\mu$ of Eqn.~\/(\ref{vectorfield1}) in powers of
$1/r$ we find
\begin{equation}
A_\mu = - 2i\bar{\sigma}_{\mu\nu}\left[\frac{1}{r} \hat{x}_\nu +
\frac{1}{r^2}\hat{n}_\nu^{(X)} \right] +
\mathcal{O}(\frac{1}{r^3}).
\end{equation}
If we define $U \in$ SU(2) as
\begin{equation}
\label{firstgaugetrafo} U = \hat{x}_\nu\alpha_\nu
\end{equation}
we have
\begin{equation}
U^{-1} \partial_\mu U = \frac{-2i
\bar{\sigma}_{\mu\nu}\hat{x}_\nu}{r},
\end{equation}
where $U^{-1}$ is given by $U^{-1} = \hat{x}_\nu\bar{\alpha}_\nu$.
Thus gauge transforming $A_\mu$ with $U$ as defined in
Eqn.~\/(\ref{firstgaugetrafo}) cancels out the $1/r$ contribution
to $A_\mu$. The additional gauge transformation required to cancel
out the $1/r^2$ contribution can be found as follows. We shall
write $U_{\rm{add}} = \mathbbm{1}_2 + \epsilon$ for the
additionally required SU(2) gauge transformation, where we assume
$\epsilon$ to be of $\mathcal{O}(r^{-1})$. The inverse gauge
transformation is $U^{-1}_{\rm{add}} = \mathbbm{1}_2
- \epsilon$. Gauge transforming again we have
\begin{equation}
A'_\mu \longrightarrow A^{''}_\mu = A'_\mu + \epsilon A'_\mu -
A'_\mu \epsilon + \epsilon  A'_\mu  \epsilon - \partial_\mu
\epsilon (\mathbbm{1}_2 - \epsilon),
\end{equation}
where $A'_\mu$ denotes the gauge potential gauge transformed by
$U$. Thus, we can read off, that the $\mathcal{O}(r^{-2})$
contribution to $A'_\mu$ will be cancelled by
$\partial_\mu\epsilon$. Hence, if we are able to write the
$\mathcal{O}(r^{-2})$ contribution to $A'_\mu$ as the derivative
of some function $\epsilon$ we will have obtained the required
additional gauge transformation. Using the fact that
\begin{equation}
\delta_{\mu\nu} = \frac{1}{2} (\alpha_\mu \bar{\alpha}_\nu +
\alpha_\nu \bar{\alpha}_\mu)
\end{equation}
we find for the $\mathcal{O}(r^{-2})$ contribution to $A'_\mu$
\begin{eqnarray}
 - 2i \, U \bar{\sigma}_{\mu\nu}U^{-1} \, \frac{\hat{n}_\nu^{(X)}}{r^2}
 & = & \frac{1}{r^2}\left[- \hat{n}_\mu^{(X)} + \mathbf{X}.\mbox{\boldmath$\alpha$}(2\hat{x}_\mu\hat{x}_\nu\bar{\alpha}_\nu
  - \bar{\alpha}_\mu)  \right] \nonumber \\
  {} & = & \partial_\mu \left[\frac{\mathbf{\hat{x}}.\mathbf{X} - (\mathbf{X}.\mbox{\boldmath$\alpha$})(\mathbf{\hat{x}}.\bar{\mbox{\boldmath$\alpha$}})}{r} \right].
\end{eqnarray}
Thus we find that the additional gauge transformation is given by
\begin{equation}
U_{\rm{add}} = \mathbbm{1}_2 + \frac{\mathbf{\hat{x}}.\mathbf{X} -
(\mathbf{X}.\alpha)(\mathbf{\hat{x}}.\bar{\alpha})}{r}.
\end{equation}
Defining the self-dual symbols $\sigma_{\mu\nu}$ analogously to
the anti-self-dual symbols $\bar{\sigma}_{\mu\nu}$
\begin{equation}
\sigma_{\mu\nu} = \frac{1}{4i}\left( \alpha_\mu\bar{\alpha}_\nu
 - \alpha_\nu \bar{\alpha}_\mu \right)
\end{equation}
it is easy to see that $U_{\rm{add}}$ is indeed an SU(2) gauge
transformation, since we may write
\begin{equation}
\mathbf{\hat{x}}.\mathbf{X} -
(\mathbf{X}.\alpha)(\mathbf{\hat{x}}.\bar{\alpha}) = - 2i
\sigma_{\mu\nu} X_\mu \hat{x}_\nu,
\end{equation}
where we have used
\begin{equation}
\alpha_\mu \bar{\alpha}_\nu = 2i \sigma_{\mu\nu} + \delta_{\mu\nu}.
\end{equation}
 Note that the $\sigma_{\mu\nu}$'s take their values in the Lie
 algebra of SU(2) as do the $\bar{\sigma}_{\mu\nu}$'s.

It is remarkable that the additional gauge transformation depends
only on the centre of mass $\mathbf{X}$ of the instanton. In the
centre of mass frame, where $\mathbf{X} = \mathbf{0}$, the
additional gauge transformation is the identity. Thus in this
frame the quadrupole term is a solution to the free equation even
without an additional gauge transformation.

The basic quantity $\hat{n}^{(k)}_\mu \bar{\alpha}_\mu$, defined
in Eqn.~\/(\ref{nhat}), gauge transformed by $U$, is given by
\begin{eqnarray}
\hat{\mathcal{N}}_s^{(k)} & \equiv & \hat{x}_\mu\alpha_\mu\;\hat{n}^{(k)}_\nu\bar{\alpha}_\nu \nonumber \\
 {} & = & 2\,\mathbf{\hat{x}}.\mathbf{x}_{(k)} -
 (\mathbf{\hat{x}}.\mbox{\boldmath$\alpha$})(\mathbf{x}_{(k)}.\bar{\mbox{\boldmath$\alpha$}})\nonumber\\
 {} & = & (\mathbf{x}_{(k)}.\mbox{\boldmath$\alpha$})(\mathbf{\hat{x}}.\bar{\mbox{\boldmath$\alpha$}}).
\end{eqnarray}
Comparing this to the singular solution discussed previously
(Eqn.~\/(\ref{matrixsingularsolution})) we find that
$\hat{\mathcal{N}}_s^{(k)}$ is precisely the matrix occurring
there, with the components of the vector $\mathbf{c}$ determined
by the components of the position parameter $\mathbf{x}_{(k)}$ of
the instanton.

We find for the dipole contribution $\psi^D_{(k)}$ to the fully
gauge transformed $\psi'_{(k)}$
\begin{equation}
\label{dipole} \psi^D_{(k)} =
\mathbf{\hat{x}}.\mbox{\boldmath$\alpha$} \, m_{(k)\mu}^D
\bar{\alpha}_\mu = - \frac{1}{r^3} \cdot \frac{2
\lambda_{(k)}^2}{\Lambda^{1/2}} \cdot
\left[\hat{\mathcal{N}}_s^{(k)} - \hat{\mathcal{N}}_s^{(X)}
\right].
\end{equation}
Note that the dipole contribution $\psi^D_{(k)}$ will be
unaffected by the additional gauge transformation $U_{\rm{add}}$.
However $\psi^D_{(k)}$ yields a contribution to the quadrupole
term, when one performs the additional gauge transformation
$U_{\rm{add}}$.

The quadrupole contribution $\psi^Q_{(k)}$ to ${\psi'}_{(k)}$,
fully gauge transformed by $U$ and $U_{\rm{add}}$, is given by
\begin{eqnarray}
\label{quadrupole} \lefteqn{\psi^Q_{(k)} =
\mathbf{\hat{x}}.\mbox{\boldmath$\alpha$} \,
m_{(k)\mu}^Q\bar{\alpha}_\mu +
\frac{\mathbf{\hat{x}}.\mathbf{X} - (\mathbf{X}.\alpha)(\mathbf{\hat{x}}.\bar{\alpha})}{r}\psi^D_{(k)} } \\
&& =  - \frac{1}{r^4} \cdot \frac{2
\lambda_{(k)}^2}{\Lambda^{1/2}} \Bigg[ 4\bigg(
\mathbf{\hat{x}}.\mathbf{X}\,\hat{\mathcal{N}}_s^{(X)} -
\mathbf{\hat{x}}.\mathbf{x}_{(k)}\hat{\mathcal{N}}_s^{(X)} +
\mathbf{\hat{x}}.\mathbf{x}_{(k)}\hat{\mathcal{N}}_s^{(k)}
\nonumber
\\
&& -\frac{1}{\Lambda}\sum_{i}\lambda_{(i)}^2
\mathbf{\hat{x}}.\mathbf{x}_{(i)}\hat{\mathcal{N}}_s^{(i)} \bigg)
+ \bigg(- \mathbf{X}^2 +
(\mathbf{X}.\alpha)(\mathbf{x}_{(k)}.\bar{\alpha})
 - \mathbf{x}_{(k)}^2 + \frac{\sum_{i}\lambda_{(i)}^2 \mathbf{x}_{(i)}^2}{\Lambda}\bigg)\Bigg] \nonumber.
\end{eqnarray}
To derive this we have used
\begin{equation}
\mathbf{X}.\mbox{\boldmath$\alpha$}\,\hat{n}^{(k)}_\mu
\bar{\alpha}_\mu = 2
\,\mathbf{\hat{x}}.\mathbf{x}_{(k)}\,\hat{\mathcal{N}}_s^{(X)} -
(\mathbf{X}.\mbox{\boldmath$\alpha$})(\mathbf{x}_{(k)}.\bar{\mbox{\boldmath$\alpha$}})
\end{equation}
and
\begin{equation}
(\mathbf{x}.\mbox{\boldmath$\alpha$})(\mathbf{x}.\bar{\mbox{\boldmath$\alpha$}})
= \mathbf{x}^2.
\end{equation}
If we write $\tilde{\psi}'_{(k)}$ for the fully gauge transformed
$\psi'_{(k)}$ we have
\begin{equation}
\label{asomptoticfield} \tilde{\psi}'_{(k)} = \psi_{(k)}^D +
\psi_{(k)}^Q + \mathcal{O}(\frac{1}{r^5}),
\end{equation}
where $\psi_{(k)}^D + \psi_{(k)}^Q$ should be a solution not only
to the Dirac equation in the instanton background but also to the
free equation. That this is indeed the case may be verified by
considering the solution Eqn.~\/(\ref{matrixsingularsolution})
with its pole shifted by $\mathbf{x}_{(j)}$ and with $\mathbf{c} = \mathbf{x}_{(i)}$, which is obviously still a solution to the free
equation. This shifted solution has the expansion
\begin{eqnarray}
\label{shifteddipole}
\lefteqn{\frac{\left[(\mathbf{x}_{(i)}.\mbox{\boldmath$\alpha$})(\mathbf{{x}}.\bar{\mbox{\boldmath$\alpha$}})
-
(\mathbf{x}_{(i)}.\mbox{\boldmath$\alpha$})(\mathbf{x}_{(j)}.\bar{\mbox{\boldmath$\alpha$}})
\right]}{(\mathbf{x} - \mathbf{x}_{(j)})^4}} \\ && = \frac{1}{r^3}
\hat{\mathcal{N}}_s^{(i)} +
\frac{1}{r^4}\left[4\,\mathbf{\hat{x}}.\mathbf{x}_{(j)}\,\hat{\mathcal{N}}_s^{(i)}-
(\mathbf{x}_{(i)}.\mbox{\boldmath$\alpha$})(\mathbf{x}_{(j)}.\bar{\mbox{\boldmath$\alpha$}})\right]
+ \mathcal{O}(\frac{1}{r^5}). \nonumber
\end{eqnarray}
Note that as usual the dipole and the quadrupole contribution are
separately solutions to the free equation. One now easily
verifies that both the dipole term $\psi_{(k)}^D$ and the
quadrupole term $\psi_{(k)}^Q$ solve the free equation, since one
finds $\psi_{(k)}^D$ to be a linear combination of terms of the
form $r^{-3}\hat{\mathcal{N}}_s^{(i)}$ and $\psi_{(k)}^Q$ to be a
linear combination of terms of the form
$r^{-4}\left[4\mathbf{\hat{x}}.\mathbf{x}_{(j)}\hat{\mathcal{N}}_s^{(i)}
-
(\mathbf{x}_{(i)}.\mbox{\boldmath$\alpha$})(\mathbf{x}_{(j)}.\bar{\mbox{\boldmath$\alpha$}})\right]$.

Thus we find that the solution to the Dirac equation in the
instanton background in this singular gauge coincides to
appropriate order with a linear combination of singular solutions
to the free equation as given by Eqn.~\/(\ref{shifteddipole}). In
this sense the instanton `washes out' the singularity of this
latter solution yielding an everywhere nicely behaved particle.

In the last section we will compare the quaternionic version of
Eqn.~\/(\ref{shifteddipole}), given by
\begin{equation}
\label{quaternionicshifteddi} \frac{q_{(i)}({\bar{q}} -
\bar{q}_{(j)})}{|q - q_{(j)}|^4} =
q_{(i)}\left[\frac{\hat{\bar{q}}}{|q|^3} +
\frac{4\,\mathbf{\hat{x}}.\mathbf{x}_{(j)}\,\hat{\bar{q}} -
\bar{q}_{(j)}}{|q|^4} \right] + \mathcal{O}(\frac{1}{|q|^5}),
\end{equation}
where $q_{(i)} = x_{(i)\mu} e_\mu$ and similarly $\hat{q} =
\hat{x}_\mu e_\mu$, as well as the asymptotic fields
$\tilde{\psi}'_{(k)}$ in quaternionic notation, to the general
quaternionic Laurent expansion. Investigating some special $N = 2$
instantons we will be able to show that the constants $b_\mu$
occurring in the quaternionic Laurent series reflect the symmetry
of the underlying instanton configuration. First we will however
briefly discuss the case of an $N = 1$ instanton.

\Section{$N= 1$ instanton in singular gauge}

In the case of an $N = 1$ instanton there exist two linearly
dependent solutions to the Weyl equation, the sum of which is
equal to zero. Thus the two solutions differ only in sign.
However, the JNR formula for an $N = 1$ instanton is largely
redundant, whereas the 't Hooft formula yields already the most
general $N = 1$ instanton. As stated previously a JNR $N =1$
instanton is related to a 't Hooft $N = 1$ instanton via a gauge
transformation. Explicitly (see \cite{Manton}), consider a JNR
instanton with potential
\begin{equation}
\rho = \frac{\lambda_{(1)}^2}{(\mathbf{x} - \mathbf{x}_{(1)})^2} +
\frac{\lambda_{(2)}^2}{(\mathbf{x} - \mathbf{x}_{(2)})^2}
\end{equation}
and a 't Hooft instanton with potential
\begin{equation}
\label{tHooftpotential} \rho_{\rm{'t Hooft}} = 1 +
\frac{\lambda_{(0)}^2}{(\mathbf{x} - \mathbf{x}_{(0)})^2}
\end{equation}
where
\begin{equation}
\mathbf{x}_{(0)} = \frac{\lambda_{(1)}^2 \mathbf{x}_{(2)} +
\lambda_{(2)}^2 \mathbf{x}_{(1)}}{\Lambda}
\end{equation}
and
\begin{equation}
\lambda_{(0)}^2 =
\frac{\lambda_{(1)}^2\lambda_{(2)}^2}{\Lambda^{2}}|\mathbf{x}_{(2)}
- \mathbf{x}_{(1)}|^2.
\end{equation}
The JNR $N =1$ instanton may then be obtained from the 't Hooft $N
= 1$ instanton by conjugating with the SU(2) matrix
\begin{equation} U_0 =
\frac{x_{(2)\mu} - x_{(1)\mu} }{|\mathbf{x}_{(2)} -
\mathbf{x}_{(1)}|}\;{\alpha}_\mu,
\end{equation}
or vice versa by conjugating with the inverse of $U_0$. This can
be verified by calculating the gauge transform of the asymptotic
field $\tilde{\psi}'_{(k)}$,  $k= 1, 2$, given by
Eqn.~\/(\ref{asomptoticfield}), with the gauge transformation
given by $U_0^{-1}$. Thus one finds the solution to the Weyl
equation in the background of the JNR $N =1$ instanton to be gauge
equivalent to the solution to the Weyl equation in the background
of the 't Hooft $N = 1$ instanton. The latter solution is given by
\begin{equation}
\label{tHooft} \psi'_{\rm{'t Hooft}} =  \pm \, 2 \lambda_{(0)}^2
\frac{(x_\mu - x_{(0)\mu})\bar{\alpha}_\mu}{|\mathbf{x} -
\mathbf{x}_{(0)}|\left((\mathbf{x} - \mathbf{x}_{(0)})^2 +
\lambda_{(0)}^2\right)^{3/2}}.
\end{equation}
However, it is a well-known fact \cite{JNR} that the JNR $N =1$
instanton and the 't Hooft $N = 1$ instanton are not only related
by a gauge transformation but also by a limiting process. Taking
the JNR scalar potential
\begin{displaymath}
\rho(\mathbf{x}) = \sum_{i = 1}^{N + 1}
\frac{\lambda_{(i)}^2}{(\mathbf{x} - \mathbf{x}_{(i)})^2}
\end{displaymath}
the 't Hooft scalar potential
\begin{displaymath}\rho(\mathbf{x}) = 1 + \sum_{i = 1}^{N}
\frac{\lambda_{(i)}^2}{(\mathbf{x} - \mathbf{x}_{(i)})^2}
\end{displaymath}
may be regained in the limit $|\mathbf{x}_{(N + 1)}| \rightarrow
\infty$, $\lambda^2_{(N + 1)} \rightarrow \infty$ with
$\lambda^2_{(N + 1)}/|\mathbf{x}_{(N + 1)}| \rightarrow 1$.
Similarly one may obtain 't Hooft's solution to the Weyl equation
in the presence of an $N = 1$ instanton, Eqn.~\/(\ref{tHooft}),
via the same limiting process from Grossman's solution, given by
Eqn.~\/(\ref{righthandedsolution}), to the Weyl equation in the
background of a JNR $N = 1$ instanton. Thus the formula
\begin{equation}
\psi^{(k)}_{\rm{'t Hooft}} = \rho_{\rm{'t Hooft}}^{1/2} \,
\partial_\mu \left(\frac{\phi_{\rm{'t Hooft}}^{(k)}}{\rho_{\rm{'t
Hooft}}}\right) \bar{\alpha}_\mu \quad k = 1, 2
\end{equation}
with
\begin{equation}
\phi^{(1)}_{\rm{'t Hooft}} = \frac{\lambda_{(0)}^2}{(\mathbf{x} -
\mathbf{x}_{(0)})^2}, \quad \phi^{(2)}_{\rm{'t Hooft}} = 1,
\end{equation}
and with $ \rho_{\rm{'t Hooft}}$ given by
Eqn.~\/{(\ref{tHooftpotential})}, will yield precisely 't Hooft's
solution as stated in Eqn.~\/(\ref{tHooft}). Expanding
Eqn.~\/{(\ref{tHooft})} we have
\begin{equation}
\psi'_{\rm{'t Hooft}} = \pm 2 \lambda_{(0)}^2 \left(
\frac{\hat{x}_\mu\bar{\alpha}_\mu}{r^3} + \frac{4
\mathbf{\hat{x}}.\mathbf{x}_{(0)} \hat{x}_\mu\bar{\alpha}_\mu -
x_{(0)\mu}\bar{\alpha}_\mu}{r^4} \right) +
\mathcal{O}(\frac{1}{r^5}).
\end{equation}
Comparing this to Eqn.~\/{(\ref{shifteddipole})}, one finds that
the expansion of 't Hooft's solution agrees with the expansion of
the singular solution with shifted pole apart from a constant
gauge transformation. Note that if one includes this rigid gauge
transformation in the solution, the 't Hooft $N = 1$ instanton is
characterised by eight rather than by five parameters. By
comparison, if we ignore an overall multiplicative real constant,
the number of parameters in Eqn.~\/{(\ref{shifteddipole})} is
seven in total, when the additional rigid gauge transformation is
included. Thus the dipole and quadrupole term together almost
determine the background $N =1$ instanton but not completely.

It will be generally convenient to include the three parameters
arising from rigid gauge transformations into the number of
parameters characterising an instanton, though they are not of
physical significance. One then finds, that the general AHDM
$N$-instanton is characterised by $8N$ parameters, the JNR $N$
instanton by $5N + 7$ and the 't Hooft $N$ instanton by $5N + 3$.

\Section{$N = 2$ instantons and interpretation of
constants}

We shall now investigate the dipole and quadrupole term of the
quaternionic Laurent expansion a bit further. As stated
previously, the dipole term of this series, given by
Eqn.~\/(\ref{singularsolution}) is characterised by one constant
$c_q$ which is the quaternionic analogue of a residue. However,
the next term of the expansion, the quadrupole term, is instead
characterised by three constants, the interpretation of which is
less clear. The quadrupole term of the Laurent series is given by
\begin{equation}
Q \equiv \sum_{i=1}^3 b_i \partial_i G(q)
\end{equation}
with the $b_i$ given by Eqn.~\/(\ref{Laurentseriesbs}).
Calculating the derivative of $G(q)$ with respect to $\partial_i$
explicitly we find
\begin{equation}
\frac{\partial}{\partial x_i} G(q) = - \left[ \frac{e_i}{|q|^4} +
\frac{4\hat{\bar{q}}}{|q|^4}\hat{x}_i \right].
\end{equation}
Notice that only derivatives with respect to the spatial
coordinates $x_i$ occur, since the derivative of $G(q)$ with
respect to the time coordinate $x_0$ is determined via the
Cauchy-Riemann-Fueter equation, once the spatial derivatives are
known. However, this explicitly breaks the symmetry between the
time coordinate and the spatial coordinates. Setting
\begin{equation}
\tilde{b}_0 e_i  - \tilde{b}_i \equiv b_i \quad \mbox{with} \quad
\tilde{b}_\mu \in \mathbbm{H}
\end{equation}
we may restore this symmetry in the equations. We then find for
the quadrupole term
\begin{eqnarray}
Q & = & \sum_{i=1}^3 - b_i \left[ \frac{e_i}{|q|^4} +
\frac{4\hat{\bar{q}}}{|q|^4}\hat{x}_i \right]\nonumber\\
& = & \frac{-\left[\tilde{b}_0 - \sum_{i=1}^3  \tilde{b}_i e_i
\right]}{|q|^4} + \frac{4\left[\sum_{\mu = 0}^3 \hat{x}_\mu
\tilde{b}_\mu\right]\hat{\bar{q}}}{|q|^4}
\end{eqnarray}
where we have used $\sum_{i=1}^3 e_i \hat{x}_i \cdot
\hat{\bar{q}}= 1 - \hat{x}_0\hat{\bar{q}}$ to derive the last
line. We may simplify notation further assuming the
$\tilde{b}_\mu$'s to be the components of some quaternionic valued
vector $\mathbf{\widetilde{b}}$ and writing $\tilde{b}_q =
\tilde{b}_0 + \sum_{i=1}^3 \tilde{b}_i e_i$ and $\tilde{\bar{b}}_q
= \tilde{b}_0 - \sum_{i=1}^3 \tilde{b}_i e_i$, respectively. We
then have
\begin{equation}
\label{LaurentseriesQ} Q = \frac{-\tilde{\bar{b}}_q}{|q|^4} +
4\,\mathbf{\hat{x}}.\mathbf{\widetilde{b}}
\frac{\hat{\bar{q}}}{|q|^4} .
\end{equation}
However, one should note that this notation is somewhat
misleading, since the components $\tilde{b}_\mu$ of $\tilde{b}_q$
are themselves quaternions. Therefore $\tilde{\bar{b}}_q$ will in
general not be the quaternionic conjugate of $\tilde{b}_q$.
Nevertheless this notation proves to be useful, as can be seen
directly when comparing Eqn.~\/(\ref{LaurentseriesQ}) to the
$\mathcal{O}(r^{-4})$ contribution to
Eqn.~\/(\ref{quaternionicshifteddi}). These two expressions have a
functionally similar form, which will allow us to read off the
constants $\tilde{b}_\mu$ for the quadrupole term of some specific
instanton configuration.

Sometimes we will however use a slightly different notation. Since
the $\tilde{b}_\mu$ are themselves quaternions we may write
\begin{equation}
\tilde{b}_\mu = \tilde{b}_{\mu\nu} e_\nu \quad {\rm{with}} \quad
\tilde{b}_{\mu\nu} \in \mathbbm{R}.
\end{equation}
We then have for Eqn.~\/(\ref{LaurentseriesQ})
\begin{equation}
Q = -\frac{\bar{e}_\mu\tilde{b}_{\mu\nu} e_\nu}{|q|^4} + 4 \,
\hat{x}_\mu \tilde{b}_{\mu\nu} e_\nu \frac{\hat{\bar{q}}}{|q|^4}.
\end{equation}
This notation will prove especially useful when investigating the
symmetry properties of the solution to the Dirac equation in the
instanton background.

Note that the number of parameters occurring in the Laurent series
up to quadrupole order is 15, ignoring an overall multiplicative
real constant, but including rigid gauge transformations. This
should be compared with 16 parameters describing an $N = 2$
instanton if one includes rigid gauge transformations. Hence, as
in the $N = 1$ instanton case, the dipole and the quadrupole term
together almost determine the background instanton but not
completely. In the remainder we will show that the parameters
$\tilde{b}_\mu$ are related to the symmetry of the underlying
two-instanton configuration.
\begin{figure}
\begin{center}
\psfrag{X1}{$\mathbf{x}_{(1)}$} \psfrag{X2}{$\mathbf{x}_{(2)}$}
\psfrag{X3}{$\mathbf{x}_{(3)}$} \psfrag{A1}{$\mathbf{a}_{(1)}$}
\psfrag{A2}{$\mathbf{a}_{(2)}$} \psfrag{A3}{$\mathbf{a}_{(3)}$}
\epsfig{file=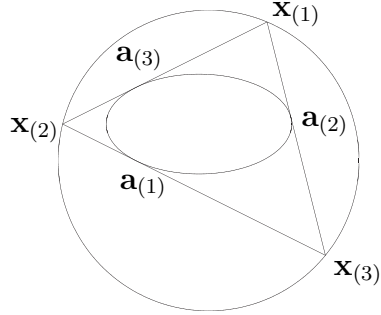}
\end{center}
\caption[A general $N = 2$ instanton with its associated Poncelet
pair of circle and ellipse.]{The circle and the ellipse associated
with a $N = 2$ instanton in $\mathbbm{R}^4$ and one member of the
porism of triangles with vertices $\mathbf{x}_{(1)}$,
$\mathbf{x}_{(2)}$ and $\mathbf{x}_{(3)}$ on the circle and
tangent to the ellipse at $\mathbf{a}_{(1)}$, $\mathbf{a}_{(2)}$
and $\mathbf{a}_{(3)}$.} \label{ellipse}
\end{figure}

With each $N = 2$ instanton in $\mathbbm{R}^4$ there is an
associated pair of a circle and an ellipse, which satisfies the
Poncelet condition, which means that there is a porism (or
one-parameter family) of triangles with vertices on the circle and
tangent to the ellipse, see Fig.~\/\ref{ellipse}. In fact, if one
such triangle exists then there is automatically a porism of
triangles and any point on the circle may be a vertex. From this
geometrical data one may reconstruct the instanton fields in JNR
form (see \cite{Manton}), by choosing one triangle of the porism
with vertices $\mathbf{x}_{(1)}$, $\mathbf{x}_{(2)}$ and
$\mathbf{x}_{(3)}$ on the circle, and tangent to the ellipse at
$\mathbf{a}_{(1)}$, $\mathbf{a}_{(2)}$ and $\mathbf{a}_{(3)}$.
Then define weights $\lambda_{(1)}^2$, $\lambda_{(2)}^2$ and
$\lambda_{(3)}^2$ up to a common multiple, by
\begin{equation}
\label{weights} \frac{\lambda_{(1)}^2}{\lambda_{(2)}^2} =
\frac{\mathbf{x}_{(1)}\,
\mathbf{a}_{(3)}}{\mathbf{a}_{(3)}\,\mathbf{x}_{(2)}}, \quad
\rm{etc.}
\end{equation}
The JNR potential is then given by
\begin{equation}
\rho = \sum_{i=1}^3 \frac{\lambda_{(i)}^2}{(\mathbf{x} -
\mathbf{x}_{(i)})^2}.
\end{equation}
If another triangle of the porism is chosen, then a different
expression for the scalar potential $\rho$ is obtained, but the
instanton will change only by a gauge transformation. This
corresponds to the gauge invariance noted by Jackiw, Nohl and
Rebbi, that moving the vertices around the circle will change the
instanton only by a gauge transformation.

\begin{figure}
\begin{center}
\psfrag{Mu}{$\mu^2$} \psfrag{Scheisse}{$\lambda^2$}
\psfrag{X1}{$\mathbf{x}_{(1)} = \mathbf{0}$}
\psfrag{X2}{$\mathbf{x}_{(2)} = - \mathbf{x}_{(3)}$}
\psfrag{X3}{$\mathbf{x}_{(3)}$} \epsfig{file=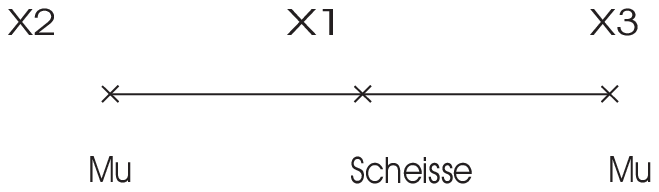}
\end{center}
\caption[$N=2$ instanton with collinear vertices.]{The three
vertices associated with an $N=2$ instanton in $\mathbbm{R}^4$ are
chosen to be collinear. This is the limiting case of a general
$N=2$ instanton in $\mathbbm{R}^4$, for which the three vertices
lie on a circle.} \label{aligned}
\end{figure}

As the first example we will consider the limiting case of an $N =
2$ instanton, where $\mathbf{x}_{(1)}$, $\mathbf{x}_{(2)}$ and
$\mathbf{x}_{(3)}$ are collinear, with $\mathbf{x}_{(1)} =
\mathbf{0}$ and $\mathbf{x}_{(2)} = - \mathbf{x}_{(3)}$, as shown
in Fig.~\/\ref{aligned}. Notice that in this limiting case the
associated ellipse becomes degenerate. The scale parameters
$\lambda_{(i)}^2$ are given by $\lambda_{(1)}^2 = \lambda^2$ and
$\lambda_{(2)}^2 = \lambda_{(3)}^2 = \mu^2$. Thus the centre of
mass is $\mathbf{X} = \mathbf{0}$, and the sum of the weights is
given by $\Lambda = \lambda^2 + 2 \mu^2$.

There are three solutions $\tilde{\psi}'_{(k)}$, whose dipole and
the quadrupole terms are given by Eqn.~\/(\ref{dipole}) and
Eqn.~\/(\ref{quadrupole}), respectively. However, since
$\tilde{\psi}'_{(1)} + \tilde{\psi}'_{(2)} + \tilde{\psi}'_{(3)} =
0$, these three solutions are linearly dependent. Thus it is
convenient to consider suitable linear combinations, which we will
choose to reflect the symmetry of the instanton. These will be
given by
\begin{equation}
\tilde{\psi}'_{(a)} = \tilde{\psi}'_{(2)} - \tilde{\psi}'_{(3)}
\quad \mbox{and} \quad \tilde{\psi}'_{(b)} = \tilde{\psi}'_{(1)}.
\end{equation}
We then find for the expansion of $\tilde{\psi}'_{(a)}$
\begin{equation}
\tilde{\psi}'_{(a)} =  -
\frac{4\mu^2}{\Lambda^{1/2}}\cdot\frac{1}{r^3}\hat{\mathcal{N}}_s^{(2)}
+ \mathcal{O}(\frac{1}{r^5}).
\end{equation}
Here we have used that since $\hat{\mathcal{N}}_s^{(k)} =
(\mathbf{x}_{(k)}.\mbox{\boldmath$\alpha$})(\mathbf{\hat{x}}.\bar{\mbox{\boldmath$\alpha$}})$
and $\mathbf{x}_{(3)} = - \mathbf{x}_{(2)}$, therefore
$\hat{\mathcal{N}}_s^{(3)} = - \hat{\mathcal{N}}_s^{(2)}$. Notice
that the quadrupole contribution to $\tilde{\psi}'_{(a)}$
vanishes. For $\tilde{\psi}'_{(b)}$ we find instead
\begin{equation}
\tilde{\psi}'_{(b)} = - \frac{1}{r^4}\cdot
\frac{2\lambda^2\mu^2}{\Lambda^{3/2}}\left[-8\mathbf{\hat{x}}.\mathbf{x}_{(2)}
\hat{\mathcal{N}}_s^{(2)} + 2\mathbf{x}_{(2)}^2 \right] +
\mathcal{O}(\frac{1}{r^5}).
\end{equation}
Here the dipole contribution is equal to zero. Notice that
$\tilde{\psi}'_{(b)}$ can be written as the sum of two shifted
dipoles, namely
\begin{equation}
\tilde{\psi}'_{(b)} \approx - \frac{2\lambda^2
\mu^2}{\Lambda^{3/2}}\left[\frac{(\mathbf{x}_{(2)}.\mbox{\boldmath$\alpha$})(\mathbf{{x}}.\bar{\mbox{\boldmath$\alpha$}}
- \mathbf{x}_{(2)}.\bar{\mbox{\boldmath$\alpha$}})}{(\mathbf{x} -
\mathbf{x}_{(2)})^4} +
\frac{(\mathbf{x}_{(3)}.\mbox{\boldmath$\alpha$})(\mathbf{{x}}.\bar{\mbox{\boldmath$\alpha$}}
- \mathbf{x}_{(3)}.\bar{\mbox{\boldmath$\alpha$}})}{(\mathbf{x} -
\mathbf{x}_{(3)})^4} \right].
\end{equation}
This will give the correct expansion of $\tilde{\psi}'_{(b)}$ at
least up to quadrupole order. In contrast, $\tilde{\psi}'_{(a)}$
can be interpreted as a pure dipole at $\mathbf{x}_{(1)} =
\mathbf{0}$.

We have for $\tilde{\psi}'_{(a)}$, $\tilde{\psi}'_{(b)}$,
respectively, in quaternionic notation
\begin{eqnarray}
\label{aligned1} \tilde{\psi}'_{(a)} & = & -
\frac{4\mu^2}{\Lambda^{1/2}} \cdot q_{(2)}
\cdot \frac{\hat{\bar{q}}}{|q|^3} \\
\label{aligned2} \tilde{\psi}'_{(b)} & = & -
\frac{4\lambda^2\mu^2}{\Lambda^{3/2}} \cdot q_{(2)} \cdot
\frac{\left[- 4\mathbf{\hat{x}}.\mathbf{x}_{(2)} \hat{\bar{q}} +
\bar{q}_{(2)} \right]}{|q|^4}.
\end{eqnarray}
Now we are able to read off the constants $\tilde{b}_\mu$
characterising the quadrupole term of the quaternionic Laurent
series. For $\tilde{\psi}'_{(b)}$,
\begin{equation}
\tilde{b}_{\mu} = \frac{4\lambda^2\mu^2}{\Lambda^{3/2}}\cdot
q_{(2)} \cdot x_{(2)\mu}
\end{equation}
or in terms of $\tilde{b}_{\mu\nu}$
\begin{equation}
\tilde{b}_{\mu\nu} = \frac{4\lambda^2\mu^2}{\Lambda^{3/2}} \cdot
x_{(2)\mu}x_{(2)\nu}.
\end{equation}
One easily checks that this gives also the correct constant
contribution to the quadrupole term. For $\tilde{\psi}'_{(a)}$ the
$\tilde{b}_\mu$ are all zero as there is no quadrupole term. The
constant $q_{(2)}$ can be interpreted as the residue of
$\tilde{\psi}'_{(a)}$. Notice that, if one chooses the vertices to
lie on some particular axis, say the $x_\mu$-axis, only one of the
parameters is not equal to zero, namely $\tilde{b}_\mu$, which is
proportional to $e_\mu$.

This nicely reflects the symmetry of the configuration of
vertices, if one assumes a generic quaternion $q= x_0 + x_i e_i$
to be invariant under the following (generalised) rotations of $\mathbbm{R}^3$.
Consider for example a rotation of $\mathbbm{R}^3$, which means
$x_i \longrightarrow R_{ij}x_j$ with $R \in$ SO(3). Note that the
quaternionic basis vectors $e_i$ are in a sense arbitrary, since
one may take linear combinations of them leading to an equivalent
description of $\mathbbm{H}$. If we identify the space spanned by
$e_i$ with the space spanned by $R^{\rm{t}}_{ji} e_i$, the map of
$\mathbbm{R}^3$ into the space of the pure quaternions with $x_0 =
0$ is spherically symmetric under the orthogonal transformation
$R$, since then $q=x_ie_i \mapsto q= R_{ij}x_j \cdot e_i= x_j
\cdot R^{\rm{t}}_{ji}e_i =x_j\tilde{e}_j \equiv x_j e_j$. In other
words, under such transformations the generic quaternion $q$ will
be invariant.

For example, the general dipole term of the Laurent expansion
$L_D$ will under a rotation $x_i \mapsto R_{ij}x_j$ with $R \in$
SO(3) transform as
\begin{equation}
L_D = (c_0 + c_i e_i) \cdot \frac{\bar{q}}{|q|^4} \; \mapsto \;
(c_0 + c_i R_{ji}\tilde{e}_j)\cdot \frac{x_0 - x_i
\tilde{e}_i}{|q|^4} \equiv (c_0  + c_i R_{ji}e_j)\cdot
\frac{\hat{\bar{q}}}{|q|^3}.
\end{equation}
If we write the product of two basic quaternions
\begin{equation}
e_i e_j = - \delta_{ij} + \epsilon_{ijk} e_k
\end{equation}
in terms of the new basis we find (since ${\rm{det}}(R) = 1$)
\begin{equation}
e_i e_j = R_{mi} \tilde{e}_m R_{nj} \tilde{e}_n = - \delta_{ij} +
{\rm{det}}(R) \, \epsilon_{ijn} R_{mn} \tilde{e}_m = - \delta_{ij}
+ \epsilon_{ijn} e_n .
\end{equation}

We are now in the position to investigate the symmetry properties
under rotations of solutions $\tilde{\psi}'_{(a)}$ and
$\tilde{\psi}'_{(b)}$, Eqn.~\/(\ref{aligned1}) and
Eqn.~\/(\ref{aligned2}), respectively. Choosing for example the
vertices in the above example to lie on the $x_0$-axis, the
background instanton is spherically symmetric in the spatial
directions. But under a spatial rotation in the above sense also
$\tilde{\psi}'_{(a)}$ remains invariant as does
$\tilde{\psi}'_{(b)}$. Thus $\tilde{\psi}'_{(a)}$ and
$\tilde{\psi}'_{(b)}$ show the same symmetry properties as the
background.

\bigskip

\begin{figure}
\begin{center}
\psfrag{a1}{$x_1$} \psfrag{a2}{$x_2$}
\psfrag{X1}{$\mathbf{x}_{(1)}, \lambda^2$}
\psfrag{X2}{$\mathbf{x}_{(2)}, \lambda^2$}
\psfrag{X3}{$\mathbf{x}_{(3)}, \lambda^2$}
\epsfig{file=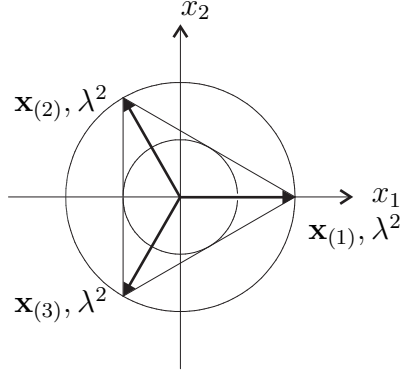}
\end{center}
\caption[Circularly symmetric $N = 2$ instanton.]{The Poncelet
pair of two concentric circles associated with the circularly
symmetric $N=2$ instanton and one member of the porism of
triangles with vertices on the outer circle and tangent to the
inner circle.} \label{circle}
\end{figure}

As the next example we will consider an instanton, where
$\mathbf{x}_{(1)}$, $\mathbf{x}_{(2)}$ and $\mathbf{x}_{(3)}$ lie
at the vertices of an equilateral triangle, such that $\sum
\mathbf{x}_{(i)} = \mathbf{0}$, $\mathbf{x}_{(i)}^2 = 1$ and where
the weights are all equal, $\lambda_{(i)}^2 = \lambda^2$. For
definiteness we shall assume that the triangle lies in the ($x_1$,
$x_2$) plane, see Fig.~\/{\ref{circle}}. We have
\begin{equation}
\label{trianglepositions} \mathbf{x}_{(1)} =
\left(\!\!\!\begin{array}{c} 0 \\ \cos{\alpha} \\ \sin{\alpha} \\
0
\end{array}\!\!\!\right), \quad
\mathbf{x}_{(2)}=\left(\!\!\!\begin{array}{c} 0\\
-\frac{1}{2}(\cos{\alpha} + \sqrt{3} \sin{\alpha}) \\
-\frac{1}{2}(\sin{\alpha} - \sqrt{3} \cos{\alpha}) \\ 0
\end{array}\!\!\!\right), \quad \mathbf{x}_{(3)}= - (\mathbf{x}_{(1)} +
\mathbf{x}_{(2)}).
\end{equation}
Here we have explicitly accounted for the fact that moving the
vertices around the circle corresponds to a gauge transformation.
Thus all configurations arising from different values of $\alpha$
in Eqn.~\/(\ref{trianglepositions}) are related via gauge
transformations. Here again we shall consider suitable linear
combinations of the three linearly dependent solutions
$\tilde{\psi}'_{(k)}$, respectively
\begin{eqnarray}
\label{circpsi1}
\tilde{\psi}'_{(1)} = & - & \frac{2\lambda}{\sqrt{3}}\bigg[\frac{q_{(1)}}{|q|^3} \\
{} & + & \frac{4}{3|q|^4}\left(-2
\mathbf{\hat{x}}.\mathbf{x}_{(2)} q_{(2)} +
\mathbf{\hat{x}}.\mathbf{x}_{(1)} q_{(1)} -
\mathbf{\hat{x}}.\mathbf{x}_{(1)}q_{(2)} -
\mathbf{\hat{x}}.\mathbf{x}_{(2)}q_{(1)} \right) \bigg]\hat{\bar{q}} + \mathcal{O}(\frac{1}{|q|^5}) \nonumber\\
\label{circpsi2} \tilde{\psi}'_{(2)} = & - &
\frac{2\lambda}{\sqrt{3}}\bigg[\frac{q_{(2)}}{|q|^3}\\
& + & \frac{4}{3|q|^4}\left(-2 \mathbf{\hat{x}}.\mathbf{x}_{(1)}
q_{(1)} + \mathbf{\hat{x}}.\mathbf{x}_{(2)} q_{(2)} -
\mathbf{\hat{x}}.\mathbf{x}_{(1)}q_{(2)} -
\mathbf{\hat{x}}.\mathbf{x}_{(2)}q_{(1)} \right) \bigg]\hat{\bar{q}} + \mathcal{O}(\frac{1}{|q|^5}) \nonumber\\
\label{circpsi3} \tilde{\psi}'_{(3)} = & - &
\frac{2\lambda}{\sqrt{3}}\bigg[-\frac{q_{(2)} + q_{(1)}}{|q|^3} \\
& + & \frac{4}{3|q|^4}\left(\mathbf{\hat{x}}.\mathbf{x}_{(1)}
q_{(1)} + \mathbf{\hat{x}}.\mathbf{x}_{(2)} q_{(2)} +
2(\mathbf{\hat{x}}.\mathbf{x}_{(1)}q_{(2)} +
\mathbf{\hat{x}}.\mathbf{x}_{(2)}q_{(1)}) \right)
\bigg]\hat{\bar{q}} + \mathcal{O}(\frac{1}{|q|^5}) \nonumber.
\end{eqnarray}
These are
\begin{eqnarray}
\label{lincom1} \psi_{(a)} & = & \tilde{\psi}'_{(1)}  +  \omega\,
\tilde{\psi}'_{(2)} + \omega^2\,
\tilde{\psi}'_{(3)} \\
\label{lincom2} \psi_{(b)} & = & \tilde{\psi}'_{(1)}  + \omega^2\,
\tilde{\psi}'_{(2)}  + \omega \,\tilde{\psi}'_{(3)}
\end{eqnarray}
with $\omega = \exp\left({\frac{2\pi k}{3}}\right) =
-\frac{1}{2}(1 - \sqrt{3}k)$. To calculate these linear
combinations it is useful to notice that
\begin{equation}
q_{(1)} = i \exp{(-k\alpha)}, \quad q_{(2)} = \omega q_{(1)},
\quad q_{(3)} = \omega^2 q_{(1)}.
\end{equation}
We then find
\begin{eqnarray}
\label{circle1} \psi_{(a)} & = & \frac{4 \sqrt{3} \lambda}{|q|^4}
\cdot \exp{( 2k\alpha)}\cdot(- i \hat{x}_1 + j
\hat{x}_2) \cdot\hat{\bar{q}} + \mathcal{O}(\frac{1}{|q|^5}) \\
\label{circle2} \psi_{(b)} & = & - \frac{2\sqrt{3}\lambda}{|q|^3}
\cdot \exp{( k\alpha)}\cdot i \cdot\hat{\bar{q}} +
\mathcal{O}(\frac{1}{|q|^4}).
\end{eqnarray}
$\psi_{(a)}$ is a pure quadrupole and $\psi_{(b)}$ is a pure
dipole with no quadrupole. The left-multiplicative factors
$\exp{(2k\alpha)}$ and $\exp{(k\alpha)}$, respectively, nicely
reflect the fact that moving the vertices around the circle
corresponds to a gauge transformation. We can now read off the
constants $\tilde{b}_\mu$ from $\psi_{(a)}$. Setting $\alpha = 0$
from now on, we have
\begin{equation}
\tilde{\mathbf{b}} = \sqrt{3}\lambda \left(\!\!\! \begin{array}{c} 0 \\ -i \\ j \\
0
\end{array}\!\!\!\right),
\end{equation}
or in terms of $\tilde{b}_{\mu\nu}$
\begin{equation}
(\tilde{b}_{\mu\nu}) = \sqrt{3} \lambda \left(%
\begin{array}{cccc}
  0 & 0 & 0 & 0 \\
  0 & -1 & 0 & 0 \\
  0 & 0 & 1 & 0\\
  0 & 0 & 0 & 0 \\
\end{array}%
\right).
\end{equation}

Since the gauge invariant data of this instanton configuration is
given by the Poncelet pair of two concentric circles
\cite{Hartshorne} as shown in Fig.~\/\ref{circle} we have circular
symmetry in the ($x_1$, $x_2$) plane. As in our previous example
the solutions $\psi_{(a)}$ and $\psi_{(b)}$,
Eqn.~\/(\ref{circle1}) and Eqn.~\/(\ref{circle2}), respectively,
exhibit the same symmetry properties as the background. Thus if
\begin{equation}
x_i \mapsto R_{ij} x_j \quad {\rm{with}} \quad R = \left(%
\begin{array}{ccc}
  \cos{\beta} & -\sin{\beta} & 0 \\
  \sin{\beta} &  \cos{\beta} & 0\\
  0 & 0 & 1
\end{array}%
\right),
\end{equation}
we find that
\begin{eqnarray}
\psi_{(a)}  & \mapsto & \exp{(-2k\beta)} \cdot \psi_{(a)} \\
\psi_{(b)}  & \mapsto & \exp{(-k\beta)} \cdot \psi_{(b)}.
\end{eqnarray}
Hence, under rotations $\psi_{(a)}$ and $\psi_{(b)}$ are invariant
up to a gauge transformation.

Note that the Weyl equation is not invariant under reflections.
Yet the background instanton is invariant under reflections in the
($x_1$, $x_2$) plane, and also under the reflection $x_3 \mapsto -
x_3$. In order to show that the spinor solutions respect this
symmetry, we will make use of the fact that two successive
reflections correspond to a rotation. We will therefore combine a
reflection, say $x_1 \mapsto -x_1$, with the reflection $x_3
\mapsto -x_3$. The combination of these is the $180^\circ$
rotation $x_i \mapsto R_{ij} x_j$, with
\begin{equation}
R = \left(%
\begin{array}{ccc}
  -1 & 0 & 0 \\
  0 & 1 & 0\\
  0 & 0 & -1
\end{array}%
\right).
\end{equation}
We find for the transformation of $\psi_{(a)}$ and $\psi_{(b)}$,
respectively,
\begin{eqnarray}
\psi_{(a)}  & \mapsto & \psi_{(a)} \\
\psi_{(b)}  & \mapsto & -\psi_{(b)}.
\end{eqnarray}
Hence, under this combination of reflections we find $\psi_{(a)}$
and $\psi_{(b)}$ to be invariant up to a gauge transformation.
Similarly, there is invariance for any reflection axis in the
($x_1$, $x_2$) plane.

\bigskip

\begin{figure}
\begin{center}
\psfrag{a1}{$x_1$} \psfrag{a2}{$x_2$}
\psfrag{X1}{$\mathbf{x}_{(1)}, \lambda^2$}
\psfrag{X2}{$\mathbf{x}_{(2)}, \mu^2$}
\psfrag{X3}{$\mathbf{x}_{(3)}, \mu^2$} \psfrag{a}{$a$}
\psfrag{b}{$b$} \epsfig{file=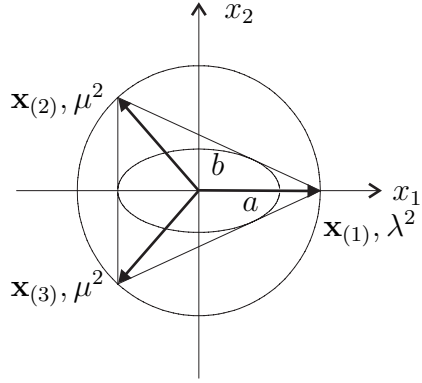}
\end{center}
\caption[$N=2$ instanton with associated concentric circle and
ellipse.]{The Poncelet pair of a concentric circle and ellipse
associated with the general $N=2$ instanton and one member of the
porism of triangles with vertices on the circle and tangent to the
ellipse.} \label{concentricellipse}
\end{figure}

So far we have only investigated two highly symmetric $N = 2$
instantons. As the final example we want to discuss the case where
the circle and ellipse, the gauge invariant data of an $N = 2$
instanton, are concentric, see Fig.~\/\ref{concentricellipse}.
Note that the moduli space associated with a general $N = 2$
instanton is --- excluding rigid gauge transformations --- a
13-dimensional manifold whereas the orbits of the conformal group
are 12-dimensional. After quotienting out the latter a general $N
= 2$ instanton is described by only one free parameter. Indeed it
may be proved that \textsl{any} $N = 2$ instanton is conformally
related to an instanton of which the gauge invariant data is given
by the Poncelet pair of a concentric circle and ellipse, the one
free parameter being the eccentricity $a/b$ of the ellipse. In
this sense an $N = 2$ instanton with associated
\textsl{concentric} circle and ellipse is the most general one.

Suppose, in Fig.~\/\ref{concentricellipse}, the radius of the
circle is $R = 1$. A porism of triangles with vertices on the
circle and tangent to the ellipse exists if and only if $a + b =
R$. For simplicity we will explicitly choose one member of the
porism of triangles, such that $\mathbf{x}_{(1)}$,
$\mathbf{x}_{(2)}$ and $\mathbf{x}_{(3)}$ are given by
\begin{equation}
\mathbf{x}_{(1)} =
\left(\!\!\!\begin{array}{c} 0 \\ 1 \\ 0 \\
0
\end{array}\!\!\!\right), \quad
\mathbf{x}_{(2)}=\left(\!\!\!\begin{array}{c} 0\\
-a \\
\sqrt{1 - a^2} \\ 0
\end{array}\!\!\!\right), \quad
\mathbf{x}_{(3)}= \left(\!\!\!\begin{array}{c} 0 \\
-a \\
-\sqrt{1 - a^2} \\
0
\end{array}\!\!\!\right).
\end{equation}
The weights associated with the vertices may be calculated using
Eqn.~\/(\ref{weights}). Scaling the weights such that
$\lambda_{(1)}^2 = \lambda^2 = 1$, we find for $\lambda_{(2)}^2$,
$\lambda_{(3)}^2$, respectively
\begin{equation}
\lambda_{(2)}^2 = \lambda_{(3)}^2 = \mu^2 = \frac{a}{b}.
\end{equation}
Again we find suitable linear combinations of the three dependent
solutions $\tilde{\psi}'_{(k)}$, $k = 1, 2, 3$, such that the two
resulting solutions form a natural basis of the solution space.
The expressions for $\tilde{\psi}'_{(k)}$ are similar to
Eqns.~\/(\ref{circpsi1}), (\ref{circpsi2}), (\ref{circpsi3}) but a
bit more elaborate. The linear combinations are
\begin{eqnarray} \label{concentric1}
\psi_{(A)} & = & \tilde{\psi}'_{(1)} + k \, \frac{1}{\sqrt{\Lambda}} (\tilde{\psi}'_{(2)}  - \tilde{\psi}'_{(3)})\\
\label{concentric2} \psi_{(B)} & = & \tilde{\psi}'_{(1)} - k \,
\frac{1}{\sqrt{\Lambda}} (\tilde{\psi}'_{(2)} -
\tilde{\psi}'_{(3)})
\end{eqnarray}
where $\Lambda = \frac{1 - a}{1 + a}$. We find that $\psi_{(A)}$
is again a pure quadrupole, and $\psi_{(B)}$ is a pure dipole.
They are, respectively
\begin{eqnarray}
\psi_{(A)} & = & \frac{1}{|q|^4} \cdot \frac{8a}{\sqrt{\Lambda}}
\cdot \left[ 4\left(-ia\hat{x}_1 + jb\hat{x}_2
\right)\cdot\hat{\bar{q}} + (b - a)\right] +
\mathcal{O}(\frac{1}{|q|^5})\\
\psi_{(B)} & = &  - \frac{1}{|q|^3} \cdot
\frac{8a}{\sqrt{\Lambda}} \cdot i \cdot \hat{\bar{q}} +
\mathcal{O}(\frac{1}{|q|^4}).
\end{eqnarray}
It is now easy to read off that the constants $\tilde{b}_\mu$
associated with $\psi_{(A)}$ are
\begin{equation}
\tilde{\mathbf{b}} = \frac{8a}{\Lambda^{1/2}}\left(\!\!\! \begin{array}{c} 0 \\ -ia \\ jb \\
0
\end{array}\!\!\!\right)
\end{equation}
and thus
\begin{equation}
(\tilde{b}_{\mu\nu}) = \frac{8a}{\Lambda^{1/2}}\left(%
\begin{array}{cccc}
  0 & 0 & 0 & 0 \\
  0 & -a & 0 & 0 \\
  0 & 0 & b & 0\\
  0 & 0 & 0 & 0 \\
\end{array}%
\right).
\end{equation}
In this example the background field exhibits just a $180^{\circ}$
rotational symmetry in the ($x_1$, $x_2$) plane and two reflection
symmetries, namely under the transformations $x_1 \mapsto -x_1$
and $x_2 \mapsto -x_2$, respectively. Under the $180^{\circ}$
rotation in the ($x_1$, $x_2$) plane we find for the
transformation of $\psi_{(A)}$ and $\psi_{(B)}$, respectively
\begin{eqnarray}
\psi_{(A)} & \mapsto & \psi_{(A)} \\
\psi_{(B)} & \mapsto & -\psi_{(B)}.
\end{eqnarray}
  Investigating the symmetry properties
of the two solutions under reflections we encounter the same
problem as in the case of the circularly symmetric instanton,
namely that the Weyl equation itself is not invariant under
reflections. However we may tackle this problem in exactly the
same way as before. We will therefore consider the transformation
of $\psi_{(A)}$ and $\psi_{(B)}$ under the following two
rotations, which are each combinations of two reflections:
\begin{equation}
\label{reflection1}
x_i \mapsto R_{ij} x_j \quad {\rm{with}} \quad R = \left(%
\begin{array}{ccc}
  -1 & 0 & 0 \\
  0 & 1 & 0\\
  0 & 0 & -1
\end{array}%
\right),
\end{equation}
and
\begin{equation}
\label{reflection2}
x_i \mapsto R_{ij} x_j \quad {\rm{with}} \quad R = \left(%
\begin{array}{ccc}
  1 & 0 & 0 \\
  0 & -1 & 0\\
  0 & 0 & -1
\end{array}%
\right).
\end{equation}
Thus, the first rotation consists of the combined reflections $x_1
\mapsto -x_1$ and $x_3 \mapsto -x_3$, whereas the second consists
of the combined reflections $x_2 \mapsto -x_2$ and $x_3 \mapsto
-x_3$. We find for the transformation of $\psi_{(A)}$ and
$\psi_{(B)}$ under the rotation given by
Eqn.~\/(\ref{reflection1})
\begin{eqnarray}
\psi_{(A)}  & \mapsto & \psi_{(A)} \\
\psi_{(B)}  & \mapsto & - \psi_{(B)}
\end{eqnarray}
and under the rotation given by Eqn.~\/(\ref{reflection2})
\begin{eqnarray}
\psi_{(A)} & \mapsto & \psi_{(A)} \\
\psi_{(B)} & \mapsto & \psi_{(B)}.
\end{eqnarray}
Note that again the two solutions respect the symmetry properties
of the background instanton.

\Section{Conclusion}

We have seen that a non-trivial solution to the free Weyl equation
in Euclidean four-space, which is bounded at infinity, necessarily
exhibits a singularity at one point at least. In the simplest case
such a solution is in quaternionic notation given by
\begin{equation}
\Psi_q = a_q \cdot \frac{{\bar{q}} - \bar{b}_q}{|q - b_q|^4}
\end{equation}
and thus parameterised by two quaternionic constants, $a_q$ and
$b_q$, respectively. Here $a_q$ is the quaternionic analogue of a
residue and as such it may be interpreted as the `spinorial
charge' of the spinor wave function. The spinor carrying this
spinorial charge is clearly localised around $b_q$, $b_q$ being
the position parameter of $\Psi_q$.

Solutions to the Weyl equation in the background of an arbitrary
JNR $N$-instanton, however, are normalisable and regular in the
whole of Euclidean four-space \cite{Grossman}. Comparing the
asymptotic behaviour of these solutions with the asymptotic
behaviour of the singular solution to the free Weyl equation we
found that the former solutions can be written up to order
$\mathcal{O}(|q|^{-4})$ as linear combinations of the latter. In
this sense the introduction of the instanton gauge field results
in a delocalisation of the spinor. Investigating some special $N =
2$ instantons we were able to show that the parameters describing
this delocalised spinor reflect the geometry of the underlying
instanton configuration.

\bigskip

\noindent \textbf{\textsc{Acknowledgements:}} \\
A.F.S. gratefully acknowledges financial support by the Gates
Cambridge Trust.

\end{document}